\documentclass[12pt]{article}
\usepackage[dvips]{graphicx}
\begin{document}
\title{\bf A case for nucleosynthesis in slowly evolving models}
\author{Geetanjali Sethi, Pranav Kumar, Sanjay Pandey 
\& Daksh Lohiya
\footnote{Inter University Centre for Astronomy and Astrophysics,
Ganeskhind, Pune, India}\\ 
Department of Physics \& Astrophysics, University of Delhi\\
Delhi 110 007, India\\
email:getsethi@physics.du.ac.in, dlohiya@iucaa.ernet.in	}
\date{}

\maketitle      
\begin{center}
Abstract\\
\end{center}
We present a case for Cosmological Nucleosynthesis in an FRW
universe in which the scale factor expands linearly with time: 
$a(t) \sim t$. It is
demonstrated that adequate amount of $^4He$ requires a baryon 
density that saturates mass bounds from galactic clusters. There
is a collataral metallicity production that is quite close to 
the lowest 
metallicity observed in metal poor Pop II stars and clouds. On 
the other hand, sites for incipient low metallicity (Pop II) star 
formation can support environments conducive to 
Deuterium production up to levels observed in the universe. 
A profile of a revised ``Standard Cosmology'' is outlined.

\section{Introduction:}

Early universe (standard big-bang) nucleosynthesis [SBBN] is
regarded as a major success for the 
Standard Big Bang [SBB] Model. As presented, SBBN results look 
rather good indeed. The observed light element abundances are 
taken to severely constrain cosmological and particle physics 
parameters. Deuterium, in particular, is regarded as
an ideal ``baryometer'' for determining the baryon content of the 
universe \cite{olive}. 
This follows from the fact that deuterium is burned away whenever
it is cycled through stars, and a belief, that there are no 
astrophysical sites (other than SBBN), capable of producing it 
in its observed abundance \cite{eps}.
The  purpose of this  article is to admit caution in adhering to 
this belief and to explore nucleosynthesis in an environment  
radically different from the Standard.

	What would be the point of such an exercise ? 

Indeed, at the outset, drastic variations from SBBN may sound 
preposterous at this time. Confidence in SBBN stems primarily 
from $D$, $^7Li$ and $^4He$ measurements.  $D$ abundance is 
measured  in the solar wind, in interstellar clouds  and, more 
recently, in the inter-galactic medium \cite{geiss,hogan}. 
The belief  that no realistic astrophysical process other than 
the Big Bang  can  produce sufficient $D$ lends 
support to its primordial origin. Further, 
$^7Li$ measurement [$^7Li/H \sim 10^{-10}$] in Pop II stars 
\cite{spite}  and the consensus \cite{yang} over the primordial 
value for the $^4He$ ratio $Y_p \ge 23.4 \%$ 
(by mass) suggest that
light element abundances are consistent with SBBN over nine 
orders of magnitude. This is  achieved by adjusting just
one parameter,  the baryon entropy ratio $\eta$. Alternative 
mechanisms for $^7Li$ production that are 
accompanied by a co-production of $^6Li$ with a later depletion 
of $^7Li$ have fallen out of favour. The debate on depletion
of $^7Li$ has been put to rest by the observation of $^6Li$ in a 
Pop II star \cite{smith}. Any depletion of $^7Li$ would have to 
be accompanied by a  complete destruction of the much more 
fragile $^6Li$. Within the SBBN scenario therefore, one seeks to 
account for the abundances of $^4He$, $D$, $^3He$ and $^7Li$ 
cosmologically, while $Be$, $B$ and $^6Li$ are generated  by 
spallation processes \cite{olive1}.

These results, however, do meet with occasional skepticism
[see eg. \cite{primak} for problems with BBN]. 
Observation of $^6Li$, for example, requires unreasonable 
suppression of astrophysical destruction of $^7Li$. On the 
other hand, the production of $^6Li$ would be accompanied by a 
simultaneous production of $^7Li$ comparable to observed levels 
\cite{rees}. This raises doubts about using observed $^7Li$ 
levels as a benchmark to evaluate SBBN. 

Further the best value of $^4He$ mass fraction, 
statistically averaged and extrapolated
to zero heavy element abundances, 
hovers around $.216 \pm .006$ for Pop II objects \cite{rana}.
Such low $^4He$ levels have also been reported     
in several metal poor HII galaxies \cite{pagel}. For example 
for  SBS 0335-052 the reported value is $Y_p = 0.21 \pm 0.01$ 
\cite{ter}. Such small values for $^4He$ would not
lead to any concordant value for $\eta$ consistent with bounds on
$^7Li$ and $D$. Of course, one could still explore a 
multi-parameter non-minimal  SBBN instead of the minimal model 
that just uses $\eta$ for a single parameter fit. Non-vanishing 
neutrino chemical potentials have been proposed to be ``natural'' 
parameters for such a venture. These conclusions have been 
criticized in \cite{yang,ter} on grounds of reliance on statistical 
over-emphasis on a few metal-poor objects with a high enough 
$^4He$ abundance to save minimal SBBN. On the other hand, there 
are objects reported with abysmally low $^4He$ levels. This is 
alarming for  minimal SBBN. For example, levels of $^4He$ 
inferred for $\mu$ Cassiopeiae A \cite{ter} and from the emission
lines of several quasars \cite{peim} are as low as 5\% and 
10 - 15\% respectively. Such low levels would most definitely 
rule out SBBN. At present one excludes such objects from SBBN 
considerations on grounds of ``our lack of understanding''of the 
environments local to these objects. As a matter
of fact,  one has to resort to specially contrived explanations to
account for low $Y_p$ values in quasars. Considering that a host 
of mechanisms for light element synthesis are discarded on grounds
of requirement of special ``unnatural'' circumstances \cite{eps}, 
it does not augur to have to resort to special explanations to 
contend with low $^4He$ emission spectra. This comment ought to 
be considered in the light of much emphasis that is 
laid on emission lines from nebulae with low metal content 
\cite{yang}. Quasars most certainly qualify for such candidates. 
Instead, one merely seems to  concentrate on classes of Pop II 
objects and HII galaxies
that would oblige SBBN. Until the dependence of light 
element abundance on sample and statistics is gotten rid  of and 
/ or fully understood, one must not close one's eyes to 
alternatives.

We end our overview of the status of SBBN  with a few comments.
Firstly the low metallicity that one sees in type II stars and 
interstellar clouds poses a problem in SBBN. There is
no object in the universe that has low abundance 
[metallicity] of heavier elements as is produced 
in SBB. One relies on some kind of re-processing, much later in 
the history of the universe, to get the low observed  
metallicity in,  for example, old clusters and 
inter-stellar clouds. This could be in the form of a generation 
of very short-lived type III stars. Such a generation of stars
may also be necessary to ionize the intergalactic medium. 
The extrapolation of $^4He$ 
abundance in type II objects and low metal (HII) galaxies, 
to its zero heavy metal abundance 
limit, presupposes that reprocessing and production of heavy
elements in type III stars is not accompanied by a significant 
change in the $^4He$ levels. A violation of this assumption,
i.e. a minute increase in $^4He$ during reprocessing (even as low
as 1 - 2 \%) would rule out the minimal SBBN. As a matter of fact, 
it is possible to account for the entire pre-galactic $^4He$ 
by such objects \cite{wagoner}.

Finally, of late \cite{lem}, the need for a careful scrutiny and 
a possible revision of the status of SBBN has also been suggested 
from  the reported high abundance of $D$ in several $Ly_\alpha$ 
systems. It may be difficult to accommodate such high abundances
within the minimal SBBN. Though the status of these observations 
is still a matter of debate, and (assuming their confirmation) 
attempts to reconcile the cosmological abundance of deuterium 
and the number of 
neutrino generations within the framework of SBB are still on, 
a reconsideration of alternate routes to deuterium in a slow
expanding universe as described in this article
could well be worth the effort. This is specially in consideration
of the stranglehold that Deuterium has on SBB in constraining the 
baryon density upper limit to not more than some 3 to 4 \% . This
constraint has been used in SBB to make out a strong case for non -
baryonic dark matter to make up the mass estimates at galactic and
cluster scales. Relying on Deuterium that is so local environment
sensitive, to predict the nature of CDM runs the risk of ``building
a colossus on a few feet of clay''\cite{borner}.

This article reports our study of nucleosynthesis in a universe 
in which the scale factor evolves linearly with time independent 
of the equation of state of matter. A strictly linear evolution 
of the cosmological scale factor
is surprisingly an excellent fit to a host of cosmological 
observations. Any model that can support such a coasting presents 
itself as a falsifiable model as far as classical cosmological 
tests are concerned as it exhibits distinguishable and verifiable 
features. 

Large scale homogeneity and isotropy observed in the universe is
incorporated in the Friedman-Robertson-Walker (FRW) metric:
\begin{equation}
\label{1}
ds^2 = dt^2 - a(t)^2[{dr^2\over {1 - Kr^2}} + r^2(d\theta^2 
+ sin^2\theta d\phi^2)]
\end{equation}
Here $K = \pm 1, 0$ is the curvature constant. In the following 
section, we summarize the 
concordance of observations in a $K = -1$ FRW cosmology in which 
the scale factor evolves linearly 
with time: $a(t) \propto t$, right from the creation event itself. 
The motivation for such an endeavor has been discussed at 
length in a series of earlier articles \cite{meetu,annu}. 
For the purpose of this article, we shall take as a conjecture
that a strictly homogeneous background FRW model coasts 
linearly with time. This can be achieved in a large class of 
non-minimally coupled theories of gravity. However, perturbations
around the homogeneous background are asumed to satisfy the 
perturbed Einstein equations. 
Section 3 describes concordant nucleosynthesis in such a model.

\section{Concordance of a linear coasting cosmology:}

{\bf \noindent Classical Cosmology tests}
\begin{itemize}
\item{\it\bf \noindent n(z), a(z):} Data on Galaxy number counts 
as a function of red-shift  along with data angular diameter 
distance as a function of red-shift do not rule out a linearly 
coasting cosmology \cite{kolb1}. However, as these tests are 
marred by evolutionary effects (and mergers), they have fallen 
into disfavour as reliable tests of a viable model.

\item{\it\bf \noindent Hubble Diagram:} With the discovery of 
Supernovae type Ia [SNe Ia] as reliable standard candles, the 
status of the Hubble test has been elevated to that of a precision
measurement. The Hubble plot relates the magnitude of a standard 
candle to its redshift in an expanding FRW universe. In 
\cite{meetu} we demonstrated how linear coasting is as 
accommodating for high red-shift objects as the standard 
non--minimal model with a small cosmological constant. 
The concordance of linear coasting with SNe1a data
finds a passing mention in the analysis of Perlmutter \cite{perl}
who noted that the Hubble plot for $\Omega_\Lambda = \Omega_M = 0$
(for which the scale factor would have a linear evolution in 
standard cosmology) is ``practically identical to the 
$\bf{best fit}$ plot for an unconstrained cosmology''. The 
concordance of this Hubble plot continues even for 
the  more recent data on SNe1a with redshifts $z \ge 1$. 
The plot almost coincides with the Hubble plot for  
$\Omega_\Lambda \approx 0.72,~ \Omega_M = 0.28$.

\item{\it\bf \noindent Age of the Universe} The age estimate of 
an ($a(t) \propto t$) universe, deduced from a measurement of the 
Hubble parameter, is given by $t_o = (H_o)^{-1}$.  The low 
red-shift SNe1a give the best value of 
$65~{\rm km~ sec^{-1}~ Mpc^{-1}}$ for the Hubble parameter.
The age of the universe turns out to be $15\times 10^9$ years.
Such an age estimate is comfortably concordant with age estimates 
of old clusters. 

\item{\it\bf \noindent Lensing Statistics:} Consistency and
concordance of linear 
coasting with gravitational lensing statistics was reported in 
\cite{abha}. 

\item{\it \bf \noindent  Density Perturbations }
\vspace{.2cm}

In \cite{pranav} we explored a conjecture that the background
universe coasts linearly with perturbations around this 
background being described by
\begin{equation}
-8\pi G \delta T^M_{\mu\nu} = \delta G_{\mu\nu}
\end{equation}
It turns out that small perturbations can 
evolve to a non-linear regime and therefore be expected to lead 
to structures at large scales. This can be seen by 
expressing the metric as
$ g_{\mu\nu} = ^{(o)} g_{\mu\nu} + \delta g_{\mu\nu}$, 
the $\delta g_{\mu\nu}$ being the perturbations. Scalar
perturbations can be decomposed in terms of eigenmodes of
the laplacian on the constant $\bar{\eta}$ surface 
(here $d\bar{\eta} = a^{-1}dt$) with eigenvalues $-k^2$
In terms of matter gauge invariant variable  $D_g$
(see eg. \cite{mukh,ruth}) and a density parameter
$$
C \equiv 4\pi G \rho_b^o a_0^2 = {3\over 2}{{8\pi G}\over {3H_o^2}}\rho_b^o 
= {3\over 2}\Omega_b  
$$
the density perturbation equation simply reduces to \cite{pranav}:
\begin{equation}
[(k^2 + 3){{e^{\bar{\eta}}}\over C} + 3]\ddot D_g +
[(k^2 + 3){{e^{\bar{\eta}}}\over C} + 2]\dot D_g
- D_gk^2  = 0.
\end{equation}
$k = 1$ corresponds to the Hubble scale which is the same as 
the curvature scale in this model. At a redshift $\approx 1000$, 
a sphere of Hubble radius subtends an
angle roughly .25 degrees. Using constraints from microwave 
background anisotropy at these angles gives $D_g \approx 10^{-5}$ 
at these scales at the last scattering surface. It is easy to see 
that modes 
$k \le 1$ do not grow. At smaller angular scales ($k >> 1$),  
the observed anisotropy is expected to fall to much lower values 
\cite{savthes}. Photon diffusion dampens anisotropies at angular 
scales smaller than about one minute. However, for such large 
values of $k$, $D_g$ has rapidly growing solutions. The 
perturbation equation becomes  
\begin{equation}
\ddot D_g + \dot D_g - Ce^{-\bar{\eta}}D_g = 0  
\end{equation}
This has exact solutions in terms of modified first and second 
type bessel functions $I_1,~K_1$:
\begin{equation}
D_g = C_1(Ce^{-\bar{\eta}})^{1\over 2}I_1((4Ce^{-\bar{\eta}})^{1\over 2})
       + C_2(Ce^{-\bar{\eta}})^{1\over 2}K_1((4Ce^{-\bar{\eta}})^{1\over 2})
\end{equation}
For large arguments, these functions have their asymptotic forms:
\begin{equation}
I_1 \longrightarrow {(Ce^{-\bar{\eta}})^{-{1\over 4}}\over {2\sqrt{\pi}}}
exp[2(Ce^{-\bar{\eta}})^{1\over 2}];~~
K_1 \longrightarrow {(Ce^{-\bar{\eta}})^{-{1\over 4}}\over {2\sqrt{\pi}}}
exp[- 2(Ce^{-\bar{\eta}})^{1\over 2}]
\end{equation}
Even if diffusion damping were to reduce the baryon density contrast to 
values as low as some $10^{-15}$, a straight forward numerical integration
of eqn(4) demonstrates that for $k \ge 3000$ the density contrast becomes 
non linear around redshift of the order 50. 

In contrast to the above, in the radiation dominated epoch,
the adiabatic approximation perturbation equations imply 
\cite{pranav}:
\begin{equation}
[(k^2 + 3){3\over {4k^2}} + {3\tilde C\over {2k^2e^{2\bar{\eta}}}}]\ddot D_g +
{3\tilde C\over {k^2e^{2\bar{\eta}}}}\dot D_g
+[{{k^2 + 3}\over 8} - {\tilde C\over {2e^{2\bar{\eta}}}}]D_g  = 0 
\end{equation}
For $\bar{\eta}$ large and negative, small $k$ pertubation equation 
reduces to 
\begin{equation}
3\ddot D_g + 6\dot D_g -k^2D_g = 0
\end{equation}
Eqns(7-8) imply that 
perturbations bounded for large negative $\eta$ damp out for 
small $k$ while large $k$ modes are oscillatory.

We conclude that fluctuations do not grow in the radiation 
dominated era, small $k$ (large scale) fluctuations do not grow 
in the matter dominated era as well. However, 
even tiny residual baryonic fluctuations $O(10^{-15})$ 
at the last scattering surface for large values of $k \ge 3000$ 
in the matter dominated era, grow to the non linear regime. 
Such a growth would be a necessary condition for structure 
formation and is not satisfied in the standard model. In the 
standard model, cold dark matter is absolutely essential 
for structure formation.

\vspace{.2cm}
\item{\it \bf \noindent The recombination epoch}
\vspace{.2cm}

        Salient features of the plasma era in a linear coasting 
cosmology have been described in \cite{astroph,savitaI,savthes}. 
Here we reproduce some of
the peculiarities of the recombination epoch. These
are deduced by making a simplifying assumption of thermodynamic 
equilibrium just before recombination. 

The probability that a photon was last scattered in the interval 
$(z,z + dz)$ can be expressed in terms of optical depth,and turns out to be:
\begin{equation}
P(z) = e^{-\tau_\gamma}{{d\tau_\gamma}\over {dz}} \approx 7.85\times 10^{-3}
({z\over {1000}})^{13.25}exp[-0.55({z\over {1000}})^{14.25}]
\end{equation}
This $P(z)$ is sharply peaked and well fitted by a gaussian of mean redshift
$z \approx 1037$ and standard deviation in redshift $\Delta z \approx 67.88$.
Thus in a linearly coasting cosmology, the last scattering surface locates at
redshift $z^* = 1037$ with thickness 
$\Delta z \approx 68$. Corresponding values in 
standard cosmology are $ z = 1065$ and $\Delta z \approx 80$.

   An important scale that determines the nature of CMBR anisotropy is the 
Hubble scale which is the same as the curvature scale for linear coasting. 
The angle subtended today, by a sphere of Hubble radius at $z^* = 1037$, 
turns out to be $\theta_H \approx 15.5$ minutes. The Hubble length determines the scale 
over which physical processes can occur coherently. Thus one expects all 
acoustic signals to be contained within an angle $
\theta_H \approx 15.5$ minutes.

We expect the nature of CMB anisotropy to follow from the above 
results. The details are still under study and shall be reported 
separately.

\vspace{.5cm}
\item{\it \bf \noindent Summary}
\vspace{.5cm}

In spite of a significantly different evolution, a linear 
coasting cosmology can not be ruled out by all the tests we have 
subjected it to so far. Linear coasting being extremely
falsifiable, it is encouraging to observe its
concordance !! In standard cosmology, falsifiability has taken 
a backstage - one just constrains the values of cosmological 
parameters subjecting the data to Bayesian statistics. Ideally, 
one would have been very content with a cosmology based on physics
tested in the laboratory. Clearly, standard cosmology does not 
pass such a test. One needs a mixture of hot and cold dark 
matter, together with (now) some form of \emph{dark energy} to 
act as a cosmological constant, to 
find any concordance with observations. In other words, one uses 
observations to parametrize theory in Standard Cosmology. 
In contrast, a universe that is born and evolves as a curvature 
dominated model has a tremendous concordance, it does not need any form of dark matter and there are 
sufficient grounds to explore models that support such a coasting.

\end{itemize}
\section{The Nucleosynthesis Constraint:}

What makes linear coasting particularly appealing is a 
straightforward adaptation of standard nucleosynthesis codes to
demonstrate that primordial nucleosynthesis is not an impediment 
for a linear coasting cosmology \cite{kapl,annu}. A linear 
evolution of the scale factor radically effects nucleosynthesis
in the early universe. With the present age of the universe some 
$15\times 10^9$ years and the $effective$ CMB temperature 2.73 K, 
the universe turns out to be some 45 years old at $10^9$ K.  
With the universe expanding at such low rates, weak interactions 
remain in equillibrium for temperature as low as $\approx 10^8$ K.
The neutron to proton ratio is determined by the n-p
mass difference and is approximately $n/p\sim exp[-15/T_9]$.
This falls to abysmally low values at temperatures below $10^9$ K.
Significant nucleosynthesis leading to helium formation commences 
only near temperatures below $\sim 5\times 10^9$K. The low n/p
ratio is not an impediment to adequate helium production. This
is because once nucleosynthesis commences, inverse beta decay 
replenishes neutrons by converting protons into neutrons and 
pumping them into the nucleosynthesis network. For baryon entropy 
ratio $\eta\approx 7.8\times 10^{-9}$, the standard 
nucleosynthesis network can be modified for linear coasting and gives $\approx 23.9\% $ Helium.
The temperatures are high enough to cause helium to burn.
Even in SBBN the temperatures are high enough for helium to burn.
However, the universe expands very rapidly in SBBN. In comparison,
the linear evolution gives enough time for successive burning 
of helium, carbon and oxygen. The metallicity yield is some $10^8$ times the metallicity 
produced in the early universe in the SBBN. The metallicity 
is expected to get distributed amongst nucleii with maximum 
binding energies per nucleon. These are nuclei with atomic 
masses between 50 and 60. This metallicity is close to
that seen in lowest metallicity objects. Figure(1) \& (2) describe 
nuclesynthesis as a function of the Baryon entropy ratio. The 
metallicity concommitantly produced with $\approx 23.9\%$ 
Helium is roughly $10^{-5}$ solar. 

The only problem that one has to contend with is the significantly
low residual deuterium in such an evolution. The desired amount
would have to be produced by the spallation processes much later 
in the history of the universe as described below.

Interestingly, the baryon entropy ratio required for the right 
amount of helium corresponds to $\Omega_b \equiv \rho_b/\rho_c = 
8\pi G \rho_b/3H_o^2 \approx 0.69$ . This closes dynamic mass 
estimates of large galaxies and clusters [see eg 
\cite{tully}]. In standard cosmology this closure is 
sought by taking recourse to non-baryonic cold dark matter. 
There is hardly any budget for non - baryonic CDM in linear 
coasting cosmology.


{\bf Deuterium Production:}

To get the observed abundances of light elements besides $^4He$,
we recall spallation mechanisms that were explored in the 
pre - 1976 days \cite{eps}. Deuterium can indeed be produced by 
the following spallation reactions:
$$
p + ^4He \longrightarrow D + ^3He; ~~ 2p \longrightarrow D + \pi^+;
$$
$$
2p \longrightarrow 2p + \pi^o,~ \pi^o \longrightarrow 2\gamma,~
\gamma +^4He \longrightarrow 2D.
$$There is no problem in producing Deuterium all the way 
to observed levels. The trouble is that under most conditions 
there is a concomitant over - production of $Li$ nuclei and 
$\gamma$ rays at unacceptable levels. Any later destruction of 
lithium in turn completely destroys $D$. As described in 
\cite{eps}, figure (3) exhibits relative production of $^7Li$  
and $D$ by spallation. It is apparent that the production of 
these nuclei to observed levels and without a collateral 
gamma ray flux is possible only if the incident (cosmic ray or any
other) beam is energized to an almost mono energetic value of 
around 400 MeV. A model that requires nearly mono energetic 
particles would be rightly considered 
$ad~hoc$ and would be hard to physically justify.

However, lithium production occurs by spallation of protons over 
heavy nuclei as well as spallation of helium over helium:
$$
p,\alpha ~+ ~C,N,O \longrightarrow Li~+~X;~~ 
p,\alpha ~+ ~Mg,Si,Fe \longrightarrow Li~+~X;~~
$$
$$
2\alpha \longrightarrow ^7Li ~+~p; ~~ \alpha ~+~D \longrightarrow p ~+~^6Li;
$$
$$
^7Be + \gamma \longrightarrow p + ^6Li; ~~ ^9Be + p \longrightarrow
\alpha +^6Li.
$$
The absence or deficiency of heavy nuclei in a target cloud and 
deficiency of alpha particles in the incident beam would clearly 
suppress lithium production. Such conditions could well have 
existed in the environments of incipient Pop II stars. 

Essential aspects of evolution of a collapsing cloud to form a low
mass Pop II star is believed to be fairly well understood 
\cite{feig,hart}. The formation
and early evolution of such stars can be discussed in terms of
gravitational and hydrodynamical processes. A protostar would 
emerge from the collapse of a molecular cloud core and would be 
surrounded by high angular momentum material forming a 
circumstellar accretion disk with bipolar outflows.
Such a star contracts slowly while the magnetic fields play a 
very important role in regulating collapse of the accretion disk 
and transferring the disk orbital angular motion to collimated 
outflows. A substantial fraction of the accreting matter is 
ejected out to contribute to the inter - stellar medium.

Empirical studies of star forming regions over the last twenty 
years have now provided direct and ample evidence for MeV 
particles produced within protostellar and T Tauri systems 
\cite{Terekhov,Torsti}. The source of such accelerated 
particle beaming is understood to be violent magnetohydrodynamic 
(MHD) reconnection events. These are analogous to solar magnetic
flaring but elevated by factors of $10^1$ to $10^6$ above levels 
seen on the contemporary sun besides being up to some 100 times 
more frequent. Accounting for characteristics in the meteoritic 
record of solar nebula from integrated effects of particle 
irradiation of the incipient sun's flaring has assumed the status 
of an industry. Protons are the primary component of particles 
beaming out from the sun in gradual flares while $^4He$ are 
suppressed by factors of ten in rapid flares to factors of a 
hundred in gradual flares\cite{Terekhov,Torsti}. Models of young 
sun visualizes it as a much larger protostar with a cooler 
surface temperature and with a very highly elevated level of 
magnetic activity in comparison to the contemporary sun. It is 
reasonable to suppose that magnetic reconnection events would 
lead to abundant release of MeV nuclei and strong shocks that 
propagate into the circumstellar matter. Considerable evidence 
for such processes in the early solar nebula has been found in 
the meteoric record. It would be fair to say that the 
hydrodynamical paradigms for understanding the earliest stages of 
stellar evolution is still not complete. However, it seems 
reasonable to conjecture that several features of collapse of a 
central core and its subsequent growth from acreting material 
would hold for low metallicity Pop II stars. Strong magnetic 
fields may well provide for a link between a central star, its 
circumstellar envelope and the acreting disk. Acceleration of 
jets of charged particles from the surface of such stars could 
well have suppressed levels of $^4He$. Such a suppression 
could be naturally expected if the particles are picked up from 
an environment cool enough to suppress ionized $^4He$ in 
comparison to ionized hydrogen. Ionized helium to hydrogen number 
ratio in a cool sunspot temperature of $\approx 3000~K$ can be 
calculated  by the Saha's ionization formula and the 
ionization energies of helium and hydrogen. This turns out to be 
$\approx~ exp(-40)$ and increases rapidly with temperature. Any 
electrodynamic process that accelerates charged particles from 
such a cool environment would yield a beam deficient in alpha 
particles. With $^4He$ content in an accelerated particle beam 
suppressed in the incident beam and with the incipient cloud 
forming a Pop II star having low metallicity in the 
target, the ``no - go'' concern of (Epstein et.al. \cite{eps}) 
is effectively circumvented. The ``no-go'' used 
$Y_\alpha /Y_p \approx .07$ in both the energetic particle
flux as well as the ambient medium besides the canonical solar 
heavy element mass fraction. Incipient Pop II environments may 
typically have heavy element fraction suppressed by more than 
five orders of magnitude while, as described above, magnetic 
field acceleration could accelerate beams of particles deficient 
in $^4He$.

One can thus have a broad energy band - all the way from a few 
MeV up to some 500 MeV per nucleon as described in the Figure (3), 
in which acceptable levels of deuterium could be ``naturally'' 
produced. The higher energy end of the band may also
not be an impediment. There are several astrophysical processes 
associated with gamma ray bursts that could produce $D$ at high 
beam energies with the surplus gamma ray flux a natural by product.

Circumventing the ``no-go'' concern of Epstein et al would be
of interest for any cosmology having an early universe 
expansion rate significantly lower than corresponding rates
for the same temperatures in early universe SBB.

{\bf Conclusions:}

Our understanding of star formation has considerably evolved 
since 1976. SBBN constraints need to be reconsidered in view of 
empirical evidence from young star forming regions. These models 
clearly imply that spallation mechanism can lead to viable and 
natural production of Deuterium and Lithium in the incipient 
environment of Pop II stars. One can conceive of cosmological 
models in which early universe nucleosynthesis produces the 
desired primordial levels of $^4He$ but virtually no $D$. Such a 
situation can arise in SBBN itself with a high baryon entropy 
ratio $\eta$. In such a universe, in principle,
Deuterium and Lithium  can be synthesized up to acceptable levels 
in the environment of incipient Pop II stars.

In SBB, hardly any metallicity is produced in the very early 
universe. Metal enrichment is supposed to be facilitated by a 
generation of Pop III stars. Pop III star formation from a 
pristine material is not well understood till date in spite of a 
lot of effort that has been expanded to that effect recently 
\cite{sneider}. It is believed that with metallicity below a 
critical transition metallicity 
($Z_{cr} \approx 10^{-4} Z_\odot$), masses of Pop III stars would 
be biased towards very high masses. Metal content higher than 
$Z_{cr}$ facilitates cooling and a formation of lower mass Pop II 
stars. In SBB, the route to Deuterium by spallation discussed in 
this article would have to follow a low metal contamination by a 
generation of Pop III stars.

Deuterium production by spallation discussed in this article 
would be good news for a host of slowly evolving cosmological 
models \cite{kapl,annu}. An FRW model with a linearly evolving 
scale factor enjoys concordance with constraints on age of the 
universe and with the Hubble data on SNe1A. Such a linear coasting
is consistent with the right amount of helium
observed in the universe and metallicity yields close to the 
lowest observed metallicities. The only problem that one has to 
contend with is the significantly low yields of deuterium in such 
a cosmology. In such a model, the first generation of stars would 
be the low mass Pop II stars and the above analysis would 
facilitate the desired deuterium yields.

In SBB, large-scale production and recycling of metals through 
exploding early generation Pop III stars leads to verifiable 
observational constraints. Such stars would be visible as 
27 - 29 magnitude stars appearing any time in every square 
arc-minute of the sky. Serious doubts have been expressed on the 
existence and detection of such signals \cite{escude}. The linear 
coasting cosmology would do away with the requirement of such 
Pop III stars altogether.

\begin{center}
\begin{figure}
\resizebox{.8\columnwidth}{!}
{\includegraphics[angle=270]{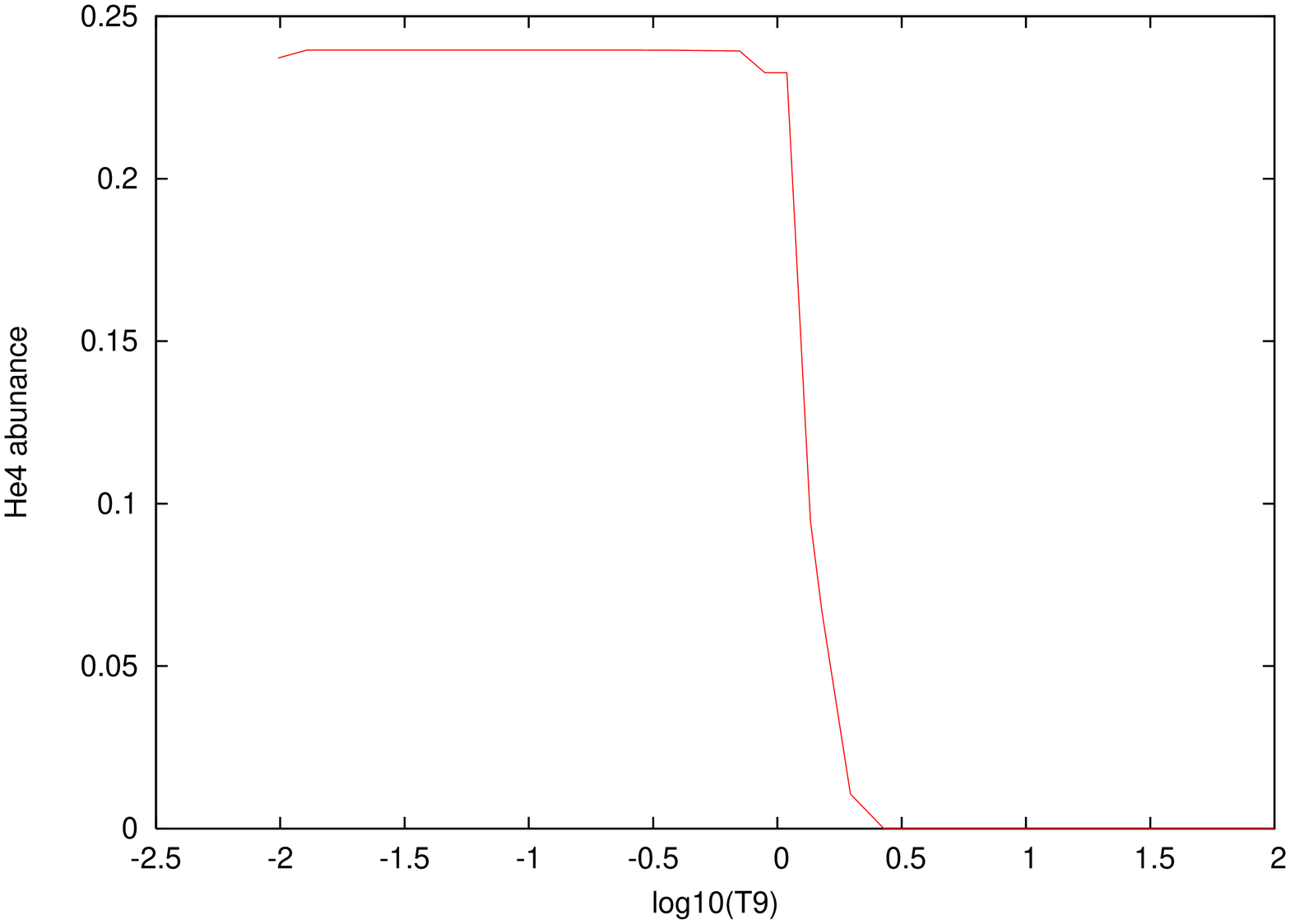}}\\
\title{Fig1(a). The figure shows abundance of He4 as a function of temperature for $\eta\approx 7.8\times 10^{-9}$. The final abundance of He4 is approximately 23 \%. It reaches this value around $T \approx T_9$ and stays same thereafter.}\\
\resizebox{.8\columnwidth}{!}
{\includegraphics[angle=270]{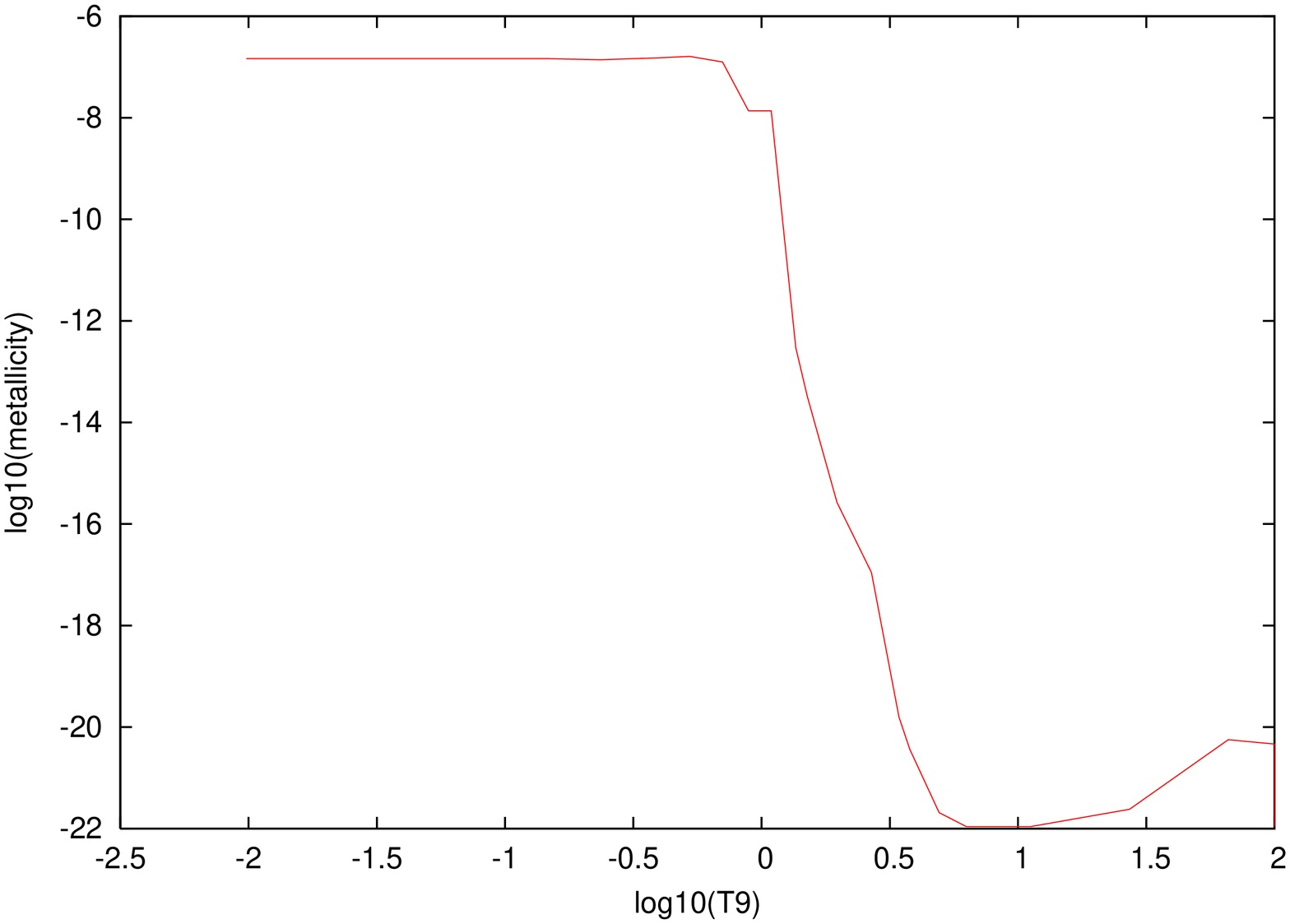}}\\
\title{Fig1(b). The figure shows metalicity as a function of temperature for $\eta\approx 7.8\times 10^{-9}$. The metallicity for a linaer coasting model is nearly equal to $ 10^{-5}$  times solar metallicity. }\\ 
\end{figure}
\end{center}
\begin{center}
\begin{figure}
\resizebox{.8\columnwidth}{!}
{\includegraphics[angle=270]{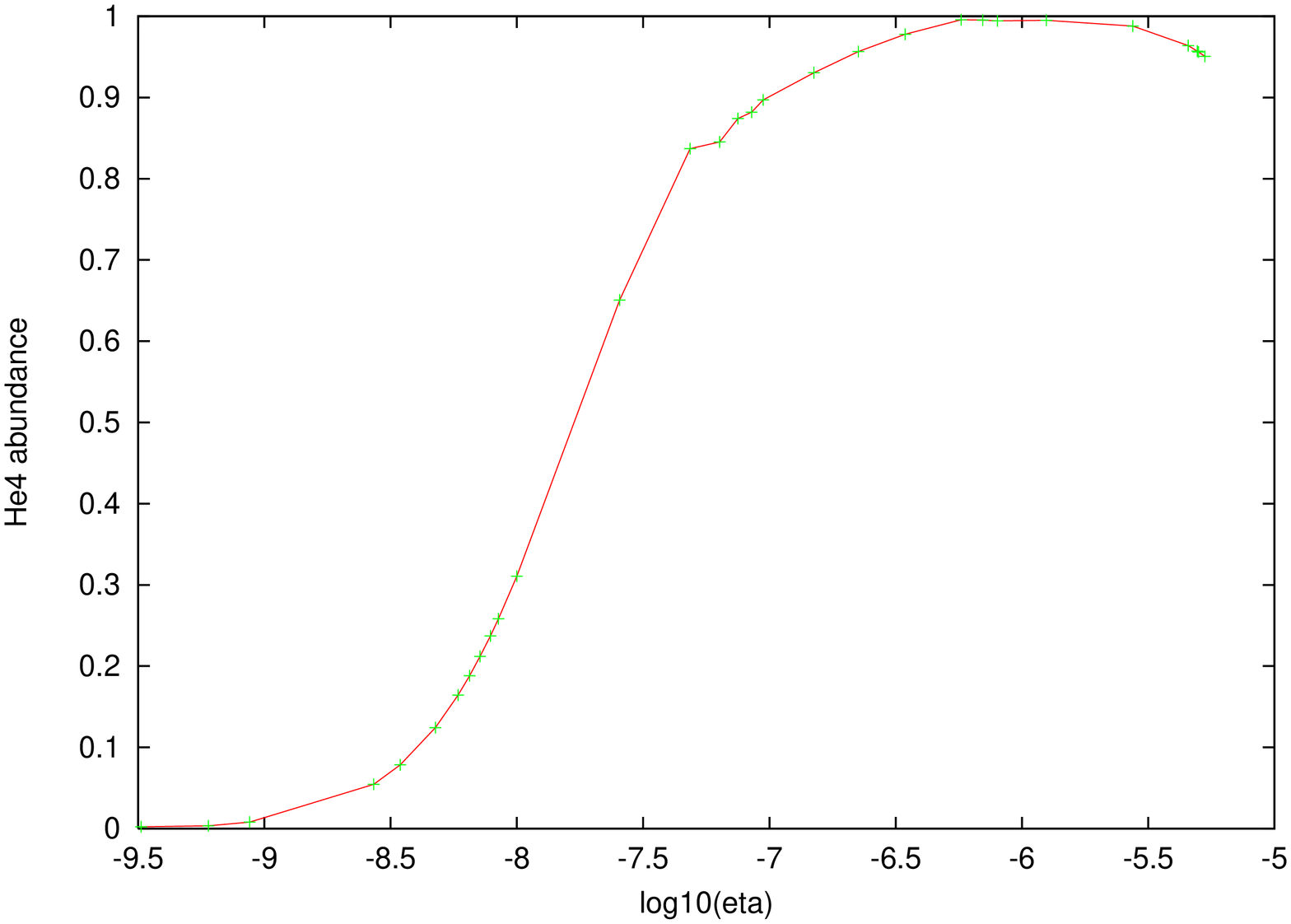}}\\
\title{Fig2(a). The figure shows He4 abundance as a function of $\eta$.}\\
\resizebox{.8\columnwidth}{!}
{\includegraphics[angle=270]{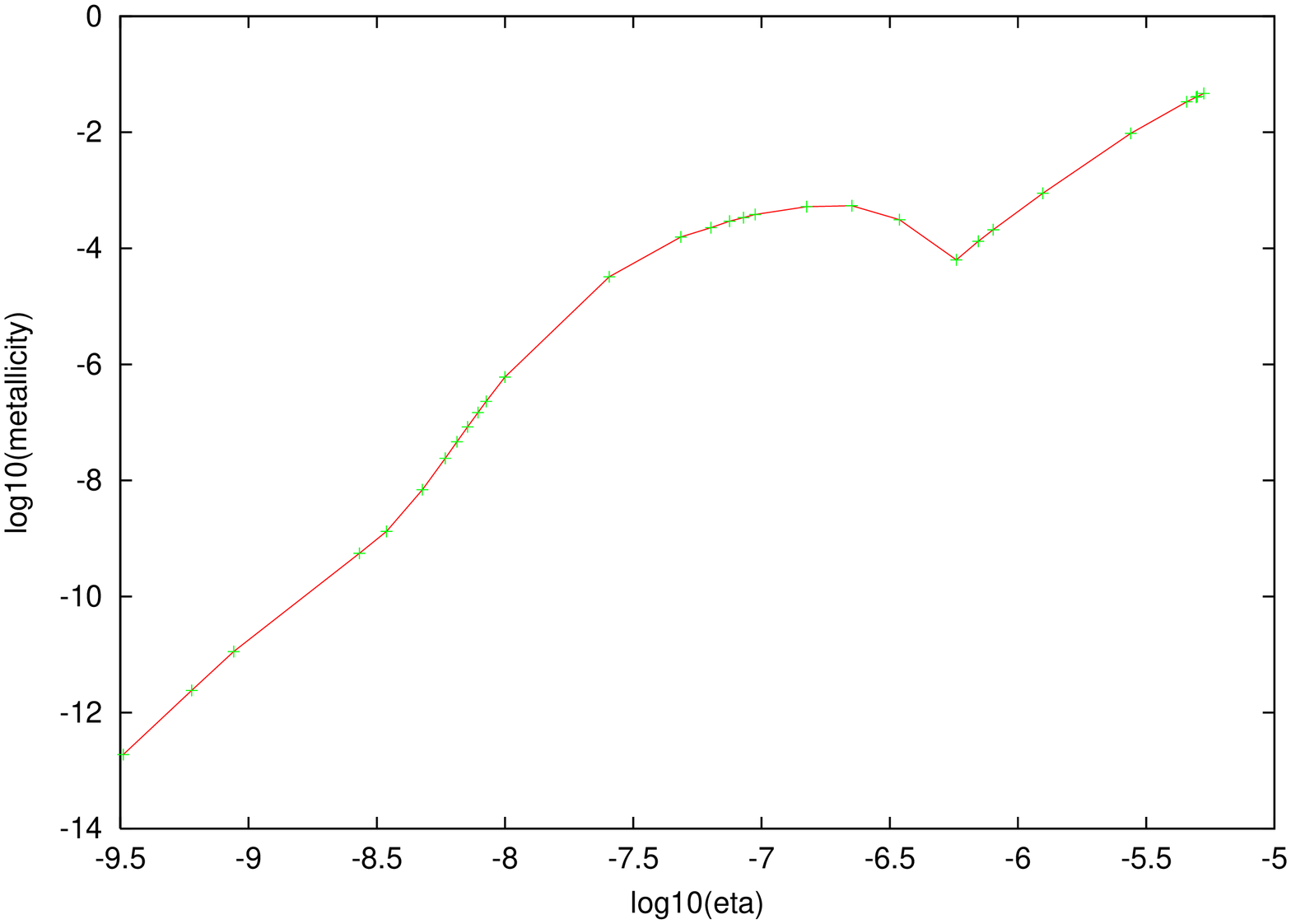}}\\
\title{Fig2(b). The figure shows metallicity as a function of $\eta$.}\\
\end{figure}
\end{center}
\eject
\begin{center}
\begin{figure}
\resizebox{.8\columnwidth}{!}
{\includegraphics{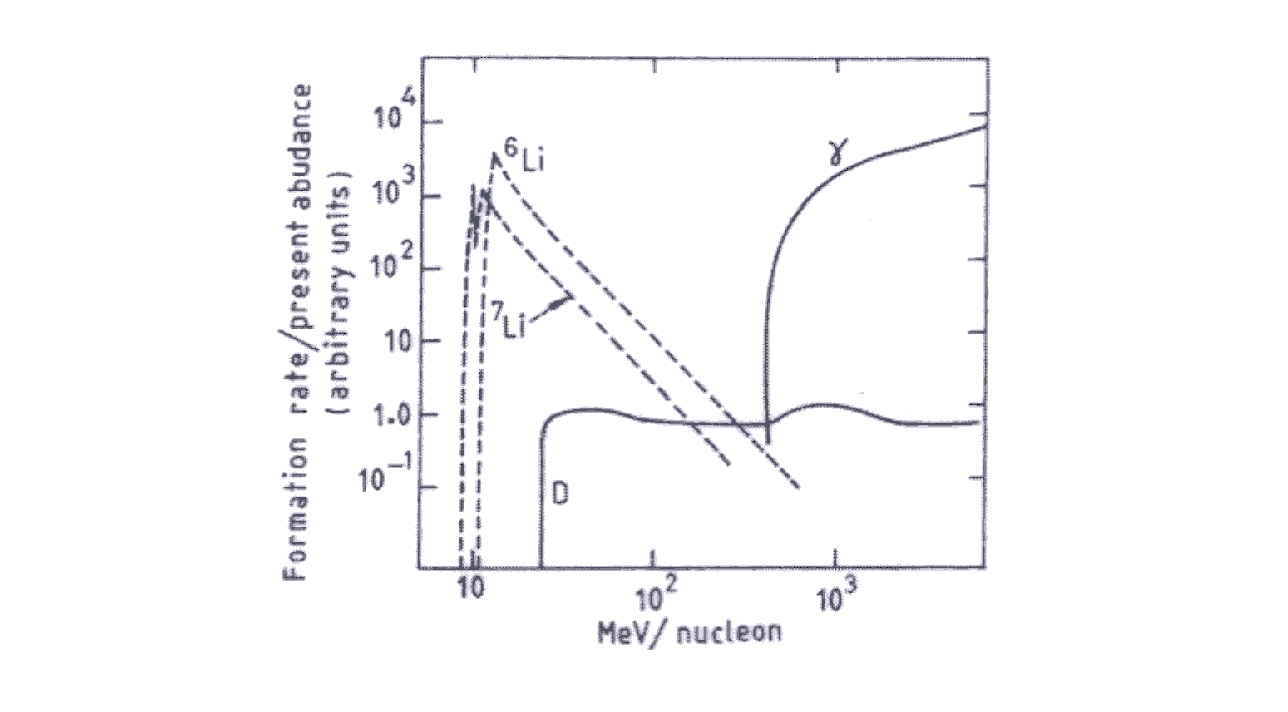}}\\
\title{ Fig3. The rates at which abundances approach their present values as a function of the energy per nucleon of the incident particle.} \cite{eps}\\
\end{figure}
\end{center}
\vskip 1cm

\vfil\eject
{\bf Acknowledgment.} \\ Daksh Lohiya and Sanjay Pandey acknowledge IUCAA for support under 
the IUCAA Associate Program. Geetanjali and Pranav ackowledge C.S.I.R for financial support. \\
\vspace{0.5cm}

\bibliography{plain}

\end{document}